\documentclass[10pt,conference]{IEEEtran}
\usepackage{fancyhdr, epsfig, epsf, amsthm, amsmath, amssymb, amsfonts, subfigure,color,dsfont}

\usepackage{cite}
\usepackage{algorithmic}
\usepackage{graphicx}
\usepackage{textcomp}
\usepackage{xcolor}
\usepackage{dblfloatfix} 
\usepackage[most]{tcolorbox}
\tcbuselibrary{breakable} 
\usepackage{enumitem}
\usepackage{url}

\def\l{\left}
\def\r{\right}
\def\({\l(}
\def\){\r)}
\def\[{\l[}
\def\]{\r]}

\IEEEoverridecommandlockouts

\renewcommand{\baselinestretch}{.82}

\newtheorem{prop}{Proposition}

\def\BibTeX{{\rm B\kern-.05em{\sc i\kern-.025em b}\kern-.08em
    T\kern-.1667em\lower.7ex\hbox{E}\kern-.125emX}}
\begin{document}
\title{Predictive Remote Control and Information Age Aware Communication Co-Design: A Gaussian Process Regression  Approach}




\title{ Predictive Control and Communication Co-Design: A Gaussian Process Regression Approach}
\author{Abanoub M. Girgis, $^\dagger$Jihong Park, Chen-Feng Liu, and Mehdi Bennis
\\[.5em]
\emph{(Invited Paper)}
    \thanks{A. M. Girgis, C.-F. Liu, and M. Bennis are with the Centre for Wireless Communications, University of Oulu, 90570 Oulu, Finland (email: \{abanoub.pipaoy,chen-feng.liu,mehdi.bennis\}@oulu.fi).}
    \thanks{$^\dagger$J. Park is with the Shool of Information Technology, Deakin University, Geelong, VIC 3220, Australia (email: jihong.park@deakin.edu.au).}}

\maketitle

\begin{abstract}
While Remote control over wireless connections is a key enabler for scalable control systems consisting of multiple actuator-sensor pairs, i.e., control systems, it entails two technical challenges. Due to the lack of wireless resources, only a limited number of control systems can be served, making the state observations outdated. Further, even after scheduling, the state observations received through wireless channels are distorted, hampering control stability. To address these issues, in this article we propose a scheduling algorithm that guarantees the age-of-information (AoI) of the last received states. Meanwhile, for non-scheduled sensor-actuator pairs, we propose a machine learning (ML) aided predictive control algorithm, in which states are predicted using a Gaussian process regression (GPR). Since the GPR prediction credibility decreases with the AoI of the input data, both predictive control and AoI-based scheduler should be co-designed. Hence, we formulate a  joint scheduling and transmission power optimization via the Lyapunov optimization framework. Numerical simulations corroborate that the proposed co-designed predictive control and AoI based scheduling achieves lower control errors, compared to a benchmark scheme using a round-robin scheduler without state prediction.
\end{abstract}
\vspace{3pt}


\section{Introduction}
\label{Sec1}

Wireless connectivity is essential in scalable control systems~\cite{shiri2019massive,liu2019taming,park2020extreme}. By decoupling the locations of sensors, actuators, and their controllers, the control system can reach its full potential in utilizing dispersed computing power and data generated at the network edge~\cite{park2019wireless}, thereby opening up many emerging use cases ranging from large-scale smart factory automation \cite{bennis2018ultrareliable} to massive drone control in real time~\cite{park2017wireless}. To achieve these goals, one key challenge is unstable and intermittent wireless connectivity that may incur communicating message distortion, hindering the stability of the control systems. A simple solution is to communicate more frequently, which is not always feasible under the limited wireless bandwidth. 

To cope with this issue, in this article we propose a machine learning (ML) based predictive control algorithm co-design with an age-of-information (AoI) aware resource allocation method. \begin{figure}[t]
    \centering
    \includegraphics[width = \columnwidth]{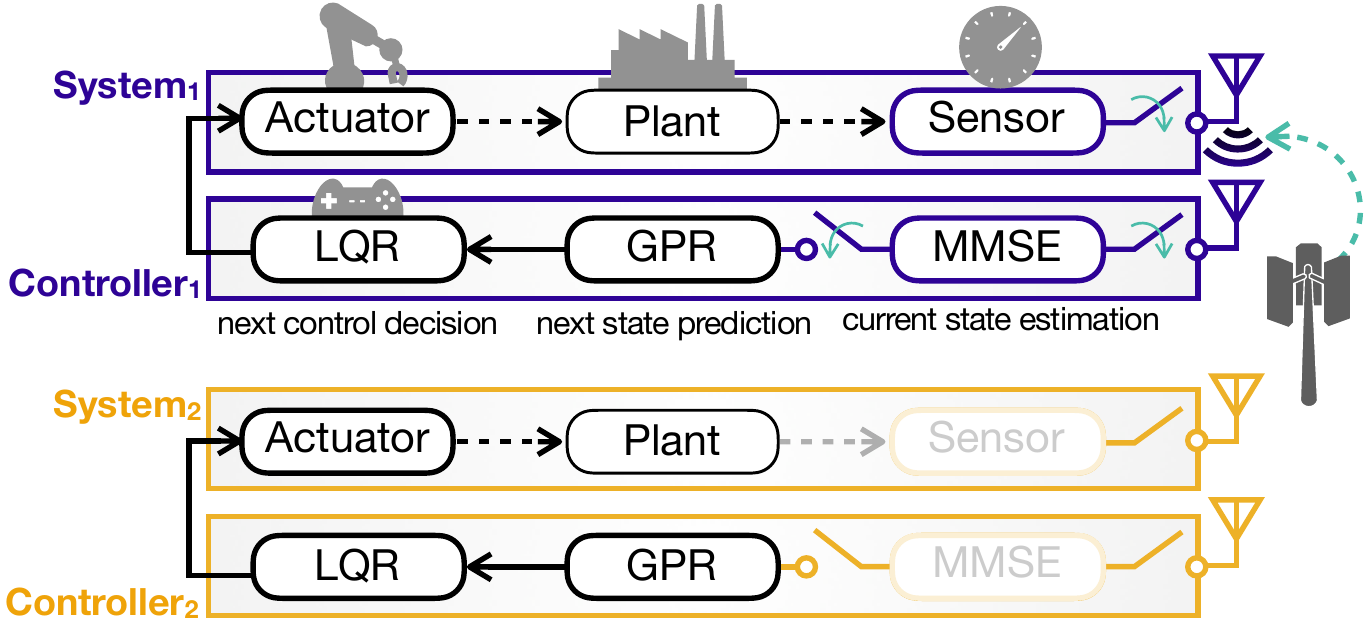}
    \caption{
        \small An illustration of $M=2$ control systems operated via both state measurement by remote sensors and state prediction by GPR.} \vspace{-5pt}
    \label{fig1}
    \end{figure}
 Fig.~\ref{fig1} illustrates the scenario under study, consisting of $M$ sensor-actuator-controller pairs, i.e., $M$ control systems. For each control system, an \emph{actuator} is controlled by a \emph{linear quadratic regulator (LQR)} whose input, i.e., future state of the {plant}, is predicted by a \emph{Gaussian process regression (GPR)} based ML method. The GPR input is the plant state history stored in a local database (DB), wherein the stored states come from (i) preceding GPR predictions and (ii) received plant state observations from a remote \emph{sensor}, if the sensor is scheduled out of $M$ control systems. Here, the control stability hinges on the GPR based state prediction credibility~\cite{roberts2013gaussian} that increases with the GPR input DB's freshness, measured by the AoI of the last received state. This AoI is dictated by the remote sensor scheduling, highlighting the importance of the communication and control co-design~\cite{zeng2019joint}. 

In this work, optimizing the scheduling of the entire systems is cast as a non-convex stochastic problem aiming at minimizing the AoI and sensor's transmission power, while guaranteeing control system stability, i.e., bounded mean-squared states. Exploiting the Lyapunov optimization framework with the drift-plus-penalty approach \cite{bennis2018ultrareliable}, the problem is recast into separable convex problems, thereby yielding closed-form optimal control solutions. In a inverted-pendulum cart control task~\cite{eisen2019control}, simulation results validate that the proposed method achieves more than $75$x less control error, measured by the mean pendulum misalignment angle, for the same amount of communication bandwidth, compared to a baseline scheme using a round-robin scheduler without state prediction.

\section{System Model}
\label{sec2}

\subsection{Control System Model}
\label{sec2_1}
Consider a wireless control system architecture as illustrated in Fig.~\ref{fig1}, consisting of a set $\mathcal{M}$ of $M$ independent linear control systems over a shared wireless channel. Each control system comprises a sensor that measures the plant's state and transmits it to the controller over a shared wireless channel. The state of the control system $i\in \mathcal{M}$ at discrete-time $k \in \mathbb{Z}_{+}$ is denoted by $\mathbf{x}_{i,k} \in \mathbb{R}^{D}$. The controller computes the appropriate control input $\mathbf{u}_{i,k} \in \mathbb{R}^{p}$ based on the received state and send actuating command to stabilize the control system. The state evolution of the control system $i$ at time $k$ is described as  \begin{equation} \small
\label{eq1}
\mathbf{x}_{i,k+1} = \mathbf{A}_{i} \mathbf{x}_{i,k} + \mathbf{B}_{i} \mathbf{u}_{i,k} + \mathbf{w}_{k},
\end{equation} where $\mathbf{A}_{i} \in \mathbb{R}^{D \times D}$ is the state transition matrix, $\mathbf{B}_{i} \in \mathbb{R}^{D \times p}$ is the control input matrix, and $\mathbf{w}_{k} \in \mathbb{R}^{D}$ is the plant noise which is independent and identically distributed Gaussian noise with zero mean and covariance matrix $W$. Here, $\mathbf{A}_{i}$ is assumed to be unstable, i.e., $\mathbf{A}_{i}$'s spectral radius $\rho(\mathbf{A}_{i}) = \text{max} \{ |\lambda_{1}(\mathbf{A}_{i})|, \cdots,| \lambda_{D}(\mathbf{A}_{i})| \} $ is larger than $1$, incurring an infinitely growing plant state over time without a proper control input. To stabilize the control system, the following three phases are considered. First, in the \textit{sensing} phase, a centralized scheduler located at the base station (BS) shared among all control systems decides which control system is scheduled to transmit and close the loop based on both the channel and control states. Then, the scheduled control system transmits its state to the controller over a wireless fading channel using analog uncoded transmission to be elaborated in sec.~\ref{sec2_2}. Next, in the \textit{processing} phase, if the control system is scheduled, the controller obtains the current plant state using a minimum mean square error (MMSE) estimator, and predicts the next plant state at use via a multi-output GPR to be discussed in Sec.~\ref{sec2_3}. Otherwise, the controller directly predicts the next plant state based on its state history using the GPR. The predicted plant state is fed to LQR generating the control input. Finally, in the \textit{actuating} phase, the controller directly applies the control input to the actuator by assuming a perfect controller-actuator channel.

\subsection{Wireless Communication model}
\label{sec2_2}

The measured control system state is transmitted over a wireless fading channel to the controller using analog uncoded transmission. Precisely, the discrete-time continuous amplitude source samples are amplified and forwarded to the controller over parallel orthogonal fading channels. The wireless communication channel is modeled as a block fading channel in which the channel is static within a time. Moreover, the channel varies independently over the times. The analog uncoded transmission has ultra-low latency compared to the conventional digital communication schemes. However, it cannot guarantee high reliability compared to the digital communication scheme, because it requires a high transmission power to ensures a specific level of signal-to-distortion ratio to enable the controller to reconstruct the transmitted signal from the control system~\cite{gastpar2003source},~\cite{hassanin2013analog}. In the uplink transmission, the received state $\mathbf{y}_{i,k} \in \mathbb{R}^{D}$ at the controller from the control system $i$ at time $k$ is written as,\begin{equation} \small 
\begin{aligned}
\label{eqt1}
\mathbf{y}_{i,k}  = \sqrt{P_{i,k}} \mathbf{C}_{i} \mathbf{H}_{i,k}  \mathbf{x}_{i,k} + \mathbf{n}_{k},
\end{aligned}
\end{equation}  where $\mathbf{x}_{i,k} = [x_{i,k}(1) \cdots x_{i,k}(D)]$ is the state transmitted by the control system $i$ at time $k$ such that $\mathbb{E} \{ |x_{i,k} (d)|^{2} \} = 1, \forall d \in \{1,\cdots,D \}$, and $P_{i,k} \in \left[ 0,P_{max} \right]$ is the transmission power of the control system $i$ at time $k$. $\mathbf{C}_{i} \in \mathbb{R}^{D \times D}$ is the state observation matrix characterized  as a full-state observation, and $\mathbf{n}_{k}$ is the additive white Gaussian noise with zero-mean and covariance matrix $ \mathbb{E}\{ \mathbf{n}_{k}^{T} \mathbf{n}_{k} \} = N_{0} \mathbf{I}_{D}$. Thus, the signal-to-noise ratio (SNR) at the controller is given as  \begin{equation} \small 
\label{eqt}
\text{SNR}_{i,k} = \frac{P_{i,k} \Vert \mathbf{H}_{i,k} \Vert^{2}}{N_{0}},
\end{equation}  where the channel state information is assumed to be perfectly known at the controller. The successful decoding of the transmitted state at the controller can be written as indicator function of a $\text{SNR}$ threshold ($\text{SNR}_{th}$) as $ \mathbb{I}_{ \{ \text{SNR}_{i,k} \geq \text{SNR}_{th}\} }$.

At each time $k$, the centralized scheduler at the BS schedules one control system depending on both the channel and control states. Let $\alpha_{i,k} \in \{0,1 \}$ be the scheduling variable of the control system $i$ at time $k$, where $\alpha_{i,k}  = 1$ when the control system $i$ is scheduled at time $k$ and $\alpha_{i,k} = 0$ otherwise.

Next, we measure the freshness of the state information at the controller using AoI, the number of elapsed time since the generation of the latest received state~\cite{kosta2017age}. In analog uncoded transmission, the AoI depends only on the time elapsed between the update generations. Hence, the control system AoI linearly increases with time if it is not scheduled or its SNR is below the threshold. Formally, the information age evolution of the control system $i$ is given as   \begin{equation} \small
\label{eq}
\beta_{i,k+1} = 1 + \left( 1 - \xi_{i,k}  \right) \beta_{i,k}, 
\end{equation} where $\beta_{i,k} \in \mathbb{Z}_{++}$ is the AoI of the control system $i$ at time $k$, and  $\xi_{i,k} = \alpha_{i,k} \mathbb{I}_{ \{ \text{SNR}_{i,k} \geq \text{SNR}_{th}\}}$ is the transmission indicator variable that depends on both the scheduling variable and the SNR indicator function. In the processing stage, there are two modes of operation for each control system depending on the transmission indicator variable. The control system $i$ at time $k$ operates in the remote-control mode if $\xi_{i,k} = 1$ and in the self-control mode otherwise. In the remote-control mode, the controller applies the MMSE estimator to filter the control state from the received signal in~\eqref{eqt1} as \begin{equation} \small
\label{eqt2}
\bar{\mathbf{x}}_{i,k}=\mathbb{E} \{ \mathbf{x}_{i,k} \vert \mathbf{y}_{i,k} \},  
\end{equation} where $\bar{\mathbf{x}}_{i,k}$ is the MMSE estimated state with estimation error $\mathbf{v}_{i,k}$ defined as a Gaussian random vector with zero-mean and covariance matrix $\mathbf{V}_{i,k} \in \mathbb{R}^{D \times D}$ given by~ \begin{equation} \small
\begin{aligned}
\label{eq}
\mathbf{V}_{i,k} &=\mathbb{E} \{ \mathbf{v}_{i,k} \mathbf{v}_{i,k}^{T} \} = \mathbb{E} \{ \left(\bar{\mathbf{x}}_{i,k} - \mathbf{x}_{i,k} \right)\left(\bar{\mathbf{x}}_{i,k} - \mathbf{x}_{i,k} \right)^{T}  \} \\ &\hspace{-15pt}= \Sigma_{x} - P^{s}_{i,k}\Sigma_{x} \mathbf{H}^{T}_{i,k} \left( P^{s}_{i,k} \mathbf{H}_{i,k} \Sigma_{x} \mathbf{H}^{T}_{i,k} + N_{0} \mathbf{I}_{D} \right)^{-1} \mathbf{H}_{i,k} \Sigma_{x} 
\end{aligned}
\end{equation} in which the control state is normally distributed with zero-mean and covariance matrix $\Sigma_{x} \in \mathbb{R}^{D \times D}$. On the contrary, the control system in the self-control mode applies the control input based on the GPR predicted state discussed in the next.

\subsection{Multi-Output Gaussian Process Regression} 
\label{sec2_3} Multi-output Gaussian process regression is a non-parametric framework for non-linear regression, where the goal is to learn the functional relationship $ \mathbf{f} \in \mathbb{R}^{D}$ between the discrete-time $t \in \mathbb{Z}_{+}$ and the MMSE estimated state $\bar{\mathbf{x}}^{s}_{i,t} \in \mathbb{R}^{D}$. This is accomplished by assuming the vector-valued functions $\bar{\mathbf{x}}^{s}_{i,t} =  \mathbf{f}(t)$, associated with different values of $t$, are random vectors and any finite number of these random vectors have a joint Gaussian distribution depending on $t$. In this way, we use the multi-output GPR to predict the state of the control system $i$ at test time $k$ from the training set of all past observed states evaluated at different times given as $\mathcal{D}_{i} = \{( t, \bar{\mathbf{x}}^{s}_{i,t} )| t = 1, \cdots, n_{d}, i=1,\cdots,M \},$ where each discrete-time $t$ is associated to $D$ outputs that represents the MMSE estimated state dimensions. Moreover, $n_{d} = \sum_{t} \xi_{i,t}$ represents the number of times the control state is sequentially observed until time $k$, hence the last time instant in which the control system transmits its state to the controller is given as $n_{d} = k - \beta_{i,k}$. In multi-output GPR, the vector-valued function $\mathbf{f} \in \mathbb{R}^{D}$ follows a Gaussian process as $ \mathbf{f} \sim \mathcal{GP} \left( \mathbf{m}, \mathbf{K}(k,k') \right)$, where $\mathbf{m} \in \mathbb{R}^{D}$ is a vector with  mean $\{ m_{d}(k)|\,d =1, \cdots, D\}$ for the outputs, which is usually taken as zero without loss of generality. The matrix $\mathbf{K}(k,k') \in \mathbb{R}^{D \times D}$ is the multi-output covariance matrix with entries are obtained through a class of multi-output kernels known as separable kernels comprising the product of a kernel on the discrete-time $k$ and a kernel representing the correlations among the outputs $f_{d}(k)$ and $f_{d'}(k')$. Throughout the paper, we consider the squared exponential (SE) kernel function for either the discrete-time or the output defined as~\cite{roberts2013gaussian},\cite{rasmussen2003gaussian},  \begin{equation}  \small
\label{eq6}
\mathcal{K}(k,k') = h^{2} \exp \left[ - \left( {k - k'}/{\lambda} \right)^{2} \right],
\end{equation} where $\lambda$ and $h$ are hyperparameters affecting the input and output scales of the SE function, respectively. For a set of observation times $N = \{1,\cdots, n_{d} \}$, the prior distribution over the vector-valued function is given by $\mathbf{f}(N) \sim \left( \mathbf{m}(N), \mathbf{K}(N,N) \right),$ where $\mathbf{m}(N) \in \mathbb{R}^{ND}$ is the mean vector. We assume that $\mathbf{f}(N)$ has zero mean, and $\mathbf{K}(N,N) \in \mathbb{R}^{ND \times ND}$ is a symmetric and block partitioned matrix with blocks $\mathbf{K}_{dd'}(N,N) = [\mathcal{K}_{dd'}(N(j),N(i))] $ for $j,i= 1, \cdots, n_{d}$ and $d,d'=1,\cdots,D$. The prior distribution of the vector-valued function at the test time $k$ is given by $\mathbf{f}(k) \sim \left( \mathbf{m}(k), \mathbf{K}(k,k) \right),$ where $\mathbf{m}(k) \in \mathbb{R}^{D}$ is the prior mean vector of the vector-valued function at the test time, and $\mathbf{K}(k,k) \in \mathbb{R}^{D \times D}$ is the covariance matrix corresponding to the covariance between the outputs at the test time. The posterior distribution of the vector-valued function $\mathbf{f}(k) = [f_{1}(k) \cdots f_{D}(k)]^{T}$  at the test time $k$ based on the training set $\mathcal{D}$ can be analytically derived as \begin{equation}  \small
\label{eq11}
\mathbf{f}(k)| \mathcal{D}, k \sim \mathcal{N} \left( \hat{\mathbf{f}}(k), \mathbf{K}^{*}_{i,k}  \right).
\end{equation} The predication mean and covariance matrix of the vector-valued function are respectively given as~\cite{alvarez2012kernels} \vspace{-5pt}  \begin{subequations}  
\small
\begin{equation}
\label{eq12a}
\hat{\mathbf{x}}_{i,k} = \hat{\mathbf{f}}(k) = \mathbf{K}(k,N) \mathbf{K}(N,N)^{-1} \mathbf{f}(N)
\end{equation}  
\begin{equation}
\begin{aligned}
\label{eq12b}
\mathbf{K}^{*}_{i,k} & = \mathbb{E}\{ \left( \hat{\mathbf{x}}_{i,k} - \mathbf{x}_{i,k} \right) \left(\hat{\mathbf{x}}_{i,k} - \mathbf{x}_{i,k} \right)^{T} \} = \mathbb{E} \{\mathbf{e}_{i,k} \mathbf{e}^{T}_{i,k}\} \\ & = \mathbf{K}(k,k) - \mathbf{K}(k,N)\mathbf{K}(N,N)^{-1} \mathbf{K}(k,N)^{T},
\end{aligned}
\end{equation}
\end{subequations} where $\mathbf{K}(k,N) \in \mathbb{R}^{D \times ND}$ is the covariance between the training and test times with blocks $  \mathbf{K}_{dd'}(k,N) = [\mathcal{K}_{dd'}(k,N(i))] $, and $\mathbf{f}(N) \in \mathbb{R}^{ND}$ in~\eqref{eq12a} is the vector-valued function that concatenates the output vectors observed over the set of observation times $N$. In addition, $\mathbf{e}^{s}_{i,k} \in \mathbb{R}^{D}$ is the state predication error defined as the difference between the true and predicated state. Based on either the MMSE estimated state or the GPR predicated state, the controller generates control input $\mathbf{u}_{i,k}$ using linear dynamic feedback control as \begin{equation} \small
\label{eqt3}
\mathbf{u}_{i,k} = - \mathbf{\Phi}_{i} (\xi_{i,k} \bar{\mathbf{x}}_{i,k}  + (1 - \xi_{i,k}) \hat{\mathbf{x}}_{i,k}),
\end{equation} where $\mathbf{\Phi} $ is the feedback gain matrix determined by LQR. In the actuating stage, the controller directly applies the control input  into the actuator to regulate the control system since the link between the controller and actuator is assumed to be occur over an ideal channel. As a result of the applied control input in~\eqref{eqt3}, the state evolution of the control system $i$ at discrete-time $k$ can be written as a switched linear time-invariant system as \begin{equation} \small
\label{eqt4}
\mathbf{x}_{i,k+1} = \left\{ \begin{array}{cc}  \mathbf{A}_{i} \mathbf{x}_{i,k} - \mathbf{B}_{i} \mathbf{\Phi}_{i} \bar{\mathbf{x}}_{i,k} + \mathbf{w}_{k}, &\text{if} \; \xi_{i,k} = 1 \\  \mathbf{A}_{i} \mathbf{x}_{i,k} - \mathbf{B}_{i} \mathbf{\Phi}_{i} \hat{\mathbf{x}}_{i,k} + \mathbf{w}_{k}, & \text{if} \; \xi_{i,k}  = 0.
\end{array}
\right.
\end{equation}

\section{Communication Control Co-design}  
\label{sec3}

To formulate the problem of joint control, communication, and GPR-based machine learning, control stability is defined as follows. The mean-square stability of the control system is given by~\cite{neely2010stability}, \vspace{-8pt}\begin{equation} \small
\label{eqt5}
\underset{K\to \infty}{\lim\sup} \frac{1}{K} \sum_{k=1}^{K}  \mathbb{E}  \{ \Vert \mathbf{x}_{i,k} \Vert^{2} \} < \infty
\end{equation} where the control system $i$ is stabilized by the controller if its long-term time-averaged mean squared state is bounded. To get the stability condition of the control system $i$ that satisfies~\eqref{eqt5}, we firstly get from~\eqref{eqt4} its state evolution as a function of both the MMSE estimation and GPR predication errors as\begin{equation} \small
\begin{aligned}
\label{eqt6}
\mathbf{x}_{i,k+1} \!=\!  \left( \mathbf{A}_{i} \!-\! \mathbf{B}_{i} \mathbf{\Phi}_{i}  \right) \mathbf{x}_{i,k} \!-\! \mathbf{B}_{i} \mathbf{\Phi}_{i} \mathbf{e}_{i,k} \!+\! \mathbf{B}_{i} \mathbf{\Phi}_{i} \xi_{i,k} \left( \mathbf{e}_{i,k} \!-\! \mathbf{v}_{i,k} \right) \!+\! \mathbf{w}_{k},
\end{aligned}
\end{equation} By taking the norm-squared and the expectation of the control state evolution in~\eqref{eqt6}, we have \begin{equation} \small
\begin{aligned} 
\label{eqt7}
\mathbb{E} \{ \Vert \mathbf{x}_{i,k+1} \Vert^{2} \} \leq & \Vert \mathbf{A}_{i} -\mathbf{B}_{i} \mathbf{\Phi}_{i} \Vert^{2} \mathbb{E} \{ \Vert \mathbf{x}_{i,k} \Vert^{2} \} - \Vert \mathbf{B}_{i} \mathbf{\Phi}_{i} \Vert^{2} \text{Tr} \{ \mathbf{K}^{*}_{i,k} \} \\&+ \xi_{i,k} \Vert \mathbf{B}_{i} \mathbf{\Phi}_{i} \Vert^{2} \left[ \text{Tr} \{ \mathbf{K}^{*}_{i,k} \} - \text{Tr} \{ \mathbf{V}_{i,k} \}  \right] + \text{Tr} \{ W \}
\end{aligned}
\end{equation} Since the pair $\left(  \mathbf{A}_{i} , \mathbf{B}_{i}  \right)$ is controllable, the  closed-loop state transition matrix $\left(\mathbf{A}_{i} - \mathbf{B}_{i} \mathbf{\Phi}_{i} \right)$ is stable matrix, thus $\Vert \mathbf{A}_{i} - \mathbf{B}_{i} \mathbf{\Phi}_{i} \Vert < 1$. By taking the long-term time-averaged on both sides of~\eqref{eqt7}, the control system $i$ satisfies the mean-square stability in~\eqref{eqt5} if and only if its arrival rate is less than its service, i.e., \begin{equation}  \small 
\label{eqt9}
\begin{aligned}
\hspace{-8pt}\underset{K\to \infty}{\lim\sup}  \frac{1}{K} \sum_{k=1}^{K} \xi_{i,k} \leq \underset{K\to \infty}{\lim\sup} \frac{1}{K} \sum_{k=1}^{K} \frac{ \text{Tr} \{ \mathbf{K}^{*}_{i,k}\} }{\left[ \text{Tr} \{ \mathbf{K}^{*}_{i,k} \} -  \text{Tr} \{ \mathbf{V}_{i,k} \}  \right]}, 
 \end{aligned} 
\end{equation} where the service rate is defined as the ratio between the GPR predication error covariance matrix trace to the difference between the GPR predication error covariance matrix trace and MMSE estimation error covariance matrix trace, and the arrival rate is defined as the transmission indicator variable. Intuitively, the AoI growing leads to the GPR predication error increasing since it depends on an outdated training set which in turn results in the control system instability. Therefore, the control system should be scheduled when the GPR predication error is greater than the MMSE estimation error to ensure the control system stability. Our primary goal is to design the communication decision variables, i.e., the transmission indicator variable and the wireless transmission power to minimize the total communication cost for each control system and to guarantee that all control systems get exactly the same level of control performance. 

\subsection{Problem Formulation}
The dynamic resource optimization can be characterized by the total communication cost of the control system, which is a weighted sum of the function of the time-averaged AoI and the function of the time-averaged transmission power for each control system and is given by \vspace{-6pt}
\begin{equation} \small
\label{eq}
C \left( \{ \bar{\beta}_{i} \}, \{ \bar{P}_{i} \} \right) = \omega_{\beta} \sum_{i=1}^{M} g_{\beta}( \bar{\beta}_{i})  +    \omega_{P} \sum_{i=1}^{M} g_{P}( \bar{P}_{i})
\end{equation} where the non-decreasing  concave functions $g_{\beta}(\beta) = \log (1+\beta)$  and  $g_{P}(P) = \log (1+P)$ are the proportionally fair cost functions of the AoI and the transmission power for each control system, respectively, the positive weights $\omega_{\beta}$ and $\omega_{P}$ indicate the relation importance of the corresponding cost functions, and the bar over the variable means its time-averaged. Then the dynamic resource optimization problem is formulated as:  \begin{subequations}\label{eqopt1} \small 
\begin{equation} \label{eqopt1_1}
 \underset{\hspace{-45pt}\mathbf{a}, \mathbf{P}}{\hspace{-50pt} (\mathcal{P}1) \quad \text{Minimize}} \quad C \left( \{  \bar{\beta}_{i} \} + \{ \bar{\hat{P}}_{i} \} \right) 
\end{equation}\vskip -15pt
 \begin{align} 
\text{subject to:}\quad & 0 \leq \alpha_{i,k} P_{i,k} \leq  P_{max},\; \forall i \in \mathcal{M}, \, \forall k    \label{eqopt1_1a} \\
& {\Vert \mathbf{H}_{i,k} \Vert^{2} \, \alpha_{i,k} \, P_{i,k}}/ {N_{0}} \geq \text{SNR}_{th}  ,\; \forall i \in \mathcal{M}, \, \forall k    \label{eqopt1_1b} \\
&  \bar{\alpha}_{i}  \leq \max \{  \bar{m}_{i,k},0 \}, \; \forall i \in \mathcal{M}   \label{eqopt1_1c}\\
&\sum_{i=1}^{M} \alpha_{i,k} \leq 1,\;  \forall k
\label{eqopt1_1d} \\
&   \alpha_{i,k} \in \{0,1\},\; \forall i \in \mathcal{M}, \, \forall k,   \label{eqopt1_1e} 
\end{align}
\end{subequations}  where $m_{i,k} =   \text{Tr} \{ \mathbf{K}^{*}_{i,k} / \}[ \text{Tr} \{ \mathbf{K}^{*}_{i,k} \} -  \text{Tr} \{ \mathbf{V}_{i,k} \}  ] $ is the upper-bound of the stability condition in~\eqref{eqt9}, and the maximum between $\bar{m}_{i,k}$ and zero in~\eqref{eqopt1_1c} is taken to ensure the constraint feasibility. Moreover, $\mathbf{a} = \{ \alpha_{i,k}: i \in \mathcal{M} \}$ is the scheduling vector, $\mathbf{P} = \{ P_{i,k}: i \in \mathcal{M} \}$ is the power allocation vector, $\hat{P}(\alpha_{i,k}) = \alpha_{i,k} P_{i,k}$ is the power allocation function that depends on the scheduling variable. The formulated dynamic resource optimization problem involves a minimizing a weighted sum of non-decreasing concave functions of the time-averaged AoI and power allocation, which is a bottleneck for the solution. To transform it into an equivalent optimization problem with a time-averaged cost function of the instantaneous AoI and power allocation instead of the cost function of the time-averaged AoI and power allocation, the non-negative auxiliary variables $\gamma^{\beta}_{i,k}$  and $\gamma^{P}_{i,k}$ are introduced using the Jensen's inequality~\cite{neely2010stability}. Then, the modified problem is given by: {\begin{subequations}\label{eqopt2} \small
\begin{equation} \label{eqopt2_1}
(\mathcal{P}2) \quad \underset{\mathbf{a}, \mathbf{P}, \mathbf{r}^{\beta}, \mathbf{r}^{P}}{\text{Minimize}} \quad \overline{ C \left( \{  \gamma^{\beta}_{i,k} \} + \{\gamma^{P}_{i,k}\} \right)} 
\end{equation}
 \begin{align}
\text{subject to:}\quad&  \bar{\beta}_{i}  \leq \bar{\gamma}^{\beta}_{i,k} ,\; \forall i \in \mathcal{M} \label{eqopt2_1a} \\  
&\bar{\hat{P}}_{i} \leq \bar{ \gamma}^{P}_{i,k}  , \; \forall i \in \mathcal{M} \label{eqopt2_1b} \\  
& 1 \leq \gamma^{\beta}_{i,k}  , \; \forall i \in \mathcal{M}, \, \forall k   \label{eqopt2_1h}\\
& 0 \leq \gamma^{P}_{i,k} \leq P_{max}, \; \forall i \in \mathcal{M}, \, \forall k   \label{eqopt2_1i} \\
&\sum_{i=1}^{M} \bar{\alpha}_{i,k} \leq 1,   \;   \forall k
\label{eqopt2_1f} \\
& ~\eqref{eqopt1_1a},~\eqref{eqopt1_1b},~\eqref{eqopt1_1c},~\eqref{eqopt1_1e}  \nonumber
\label{eqopt2_1p}
\end{align}
\end{subequations} \normalsize where \eqref{eqopt2_1h} and \eqref{eqopt2_1i} constraints are defined to bound the auxiliary variables, $\mathbf{r}_{k}^{\beta} = \{  \gamma^{\beta}_{i,k} : i \in \mathcal{M} \}$ and  $\mathbf{r}_{k}^{P} = \{  \gamma^{P}_{i,k} : i \in \mathcal{M} \}$ are the vectors of the introduced auxiliary variables. According to~\cite{neely2010stability}, the optimal solution to the transformed problem can be directly tuned into an optimal solution to the original problem. The non-convexity and stochasticity of the optimization problem $\mathcal{P}2$ is due to the constraint~\eqref{eqopt2_1f} since the control system scheduling decision depends on its own decision on all others control systems scheduling decisions. Hence, Lyapunov optimization framework is introduced to overcome this whereby, the inequality constraints in the transformed problem can be satisfied by introducing corresponding virtual queues and keeping them stable. Specifically to satisfy the auxiliary constraints in~\eqref{eqopt2_1a} and~\eqref{eqopt2_1b}, the virtual queues are introduced for the AoI and power allocation, respectively, and the dynamics of the virtual queues evolve as \vspace{-5pt}  \begin{equation}\small
\label{eqt11}
Q^{\beta}_{i,k+1} = \max \{Q^{\beta}_{i,k} - \gamma^{\beta}_{i,k}, 0  \} + \beta_{i,k}
\end{equation}   
\vspace{-4pt}  \begin{equation} \small
\label{eqt12}
Q^{P}_{i,k+1} = \max \{Q^{P}_{i,k} - \gamma^{P}_{i,k}, 0  \} + \hat{P}_{i,k},
\end{equation}   where $Q^{\beta}_{i,0} = 0$, $Q^{P}_{i,0} = 0$, $\beta_{i,k}$, and $\hat{P}_{i,k}$ will be optimized at each time $k$. Similarly, to ensure the control stability constraint in~\eqref{eqopt1_1c}, the virtual queue is introduced and the dynamics of the virtual queue evolve as \vspace{-5pt}  \begin{equation} \small 
\label{eqt13}
Q^{S}_{i,k+1} = \max \{Q^{S}_{i,k} - \max \{\bar{m}_{i,k},0 \} , 0  \} + \alpha_{i,k}, 
\end{equation} where $Q^{S}_{i,0} = 0$ and $\alpha_{i,k}$ will be optimized at each time $k$.  Intuitively, the auxiliary variables $\gamma^{\beta}_{i,k}$, $\gamma^{P}_{i,k}$, and the upper-bound of the stabity condition $\max \{\bar{m}_{i,k},0\}$ can be viewed as the service rate of the virtual queues $Q^{\beta}_{i,k}$, $Q^{P}_{i,k}$, and $Q^{S}_{i,k}$ , respectively, while $\beta_{i,k}$, $\hat{P}_{i,k}$, and $\alpha_{i,k}$ can be viewed as the arrival rate of such virtual queues. The constraints~\eqref{eqopt2_1a},~\eqref{eqopt2_1b}, and~\eqref{eqopt1_1c} are satisfied only when the virtual queues are stable. Let $\mathcal{X}_{k} = \{Q^{\beta}_{i,k},Q^{P}_{i,k}, Q^{S}_{i,k}, i \in \mathcal{M}  \}$ denote the matrix of the virtual queues then the Lyapunov function is defined as follows to represents a scalar metric of the queue congestion, \vspace{-6pt} \begin{equation} \small 
\label{eqt14}
\mathcal{L}\left( \mathcal{X}_{k} \right) = \frac{1}{2} \left[ \sum^{M}_{i=1} (Q^{\beta}_{i,k})^{2} + \sum^{M}_{i=1} (Q^{P}_{i,k})^{2} + \sum^{M}_{i=1} (Q^{S}_{i,k})^{2} \right],
\end{equation} where a small value of $\mathcal{L}\left( \mathcal{X}_{k} \right)$ implies small virtual queues and the queues have strong stability. To ensure strong stability by persistently  pushing the Lyapunov function towards lower congestion state, and to optimize the cost at the same time, the Lyaounov conditional drift-plus-penalty function is defined as~\cite{neely2010stability} \vspace{-5pt}  \begin{equation}\small 
\label{eqt15}
\Delta \left(  \mathcal{X}_{k}  \right) = \mathbb{E} \left[ \mathcal{L}\left( \mathcal{X}_{k+1} \right)  - \mathcal{L}\left( \mathcal{X}_{k} \right) + V C ( \{  \gamma^{\beta}_{i,k} \} + \{\gamma^{P}_{i,k}\} ) \right],
\end{equation}  where the control parameter $V \geq 0$ represents an importance weight on how much we emphasize the total communication cost of the control system compared to the queue stability. With the dynamics of the constructed virtual queues, the upper-bound on the drift-plus-penalty is derived as follows. At each time $k$, for any observed queue and channel states, the Lyapunov drift-plus-penalty under any control strategy satisfies the following inequality. \vspace{-6pt}  \begin{equation}
\small
\label{eqt16}
\begin{aligned}
\Delta \left(  \mathcal{X}_{k}  \right) \leq B & + \mathbb{E} \left[ \sum^{M}_{i=1} ( V \omega_{\beta} g_{\beta} (\gamma^{\beta}_{i,k}) -  Q^{\beta}_{i,k} \gamma^{\beta}_{i,k} ) \right] \\& + \mathbb{E} \left[ \sum^{M}_{i=1} ( V \omega_{P} g_{P} (\gamma^{P}_{i,k}) -  Q^{P}_{i,k} \gamma^{P}_{i,k} ) \right] \\ & + \mathbb{E} \left[ \sum^{M}_{i=1} Q^{\beta}_{i,k} \beta_{i,k} \right] + \mathbb{E} \left[ \sum^{M}_{i=1} Q^{P}_{i,k} \hat{P}_{i,k} \right] \\ & -  \mathbb{E} \left[ \sum^{M}_{i=1} Q^{S}_{i,k} (\max \{\bar{m}_{i,k,0} \} -  \alpha_{i,k}\}) \right]
\end{aligned}
\end{equation} 

According to the general Lyapunov optimization approach, instead of directly minimizing the drift-plus-penalty expression, the dynamic control congestion control strategy is obtained by minimizing the right hand side of~\eqref{eqt16} at each time, yielding:
\vspace{-2pt}  
\begin{equation} \label{eqopt3_1}
\begin{aligned}
&\hspace{22pt} (\mathcal{P}3)\quad \underset{\mathbf{a}, \mathbf{P}, \mathbf{r}^{\beta}, \mathbf{r}^{P}}{\text{Minimize}} \quad J_{i,k} \qquad \qquad \qquad \qquad \qquad \qquad  \\& \hspace{27pt}  \text{s.t}  ~\eqref{eqopt1_1a},~\eqref{eqopt1_1b},~\eqref{eqopt1_1c},~\eqref{eqopt1_1e},~\eqref{eqopt2_1h}, \; \eqref{eqopt2_1i}, \; \eqref{eqopt2_1f}, \nonumber
\end{aligned} 
\end{equation} where $J_{i,k}$ is the upper-bound on the drift-plus-penalty in~\eqref{eqt16}. We derive the closed-form solutions of this problem in the following proposition.
\begin{prop}
\label{p1}
The optimal solutions of $\mathcal{P}3$ are given as follows.

\begin{equation} \small
\label{eqt17}
\gamma^{\beta^{*}}_{i,k} = \max \left\lbrace ({V \omega_{\beta} - Q^{\beta}_{i,k}})/{Q^{\beta}_{i,k}}, 1 \right\rbrace
\end{equation}     

 \vspace{-2pt}\begin{equation} \small 
\label{eqt18}
\gamma^{P^{*}}_{i,k} = \min \left[  \max \{ ({V \omega_{P} - Q^{P}_{i,k}})/{Q^{P}_{i,k}},0 \},P_{max}   \right]
\end{equation} 

 \vspace{-2pt} \begin{equation} \small
\label{eqt19}
P^{*}_{i,k} = \left\{ \begin{array}{cc}  {\text{SNR}_{th} N_{0}}/{\Vert \mathbf{H}_{i,k} \Vert^{2}}, &\text{if} \; Q^{P}_{i,k} \geq 0 \\  P_{max}, & \text{otherwise}
\end{array}
\right.
\end{equation} 

\vspace{-4pt} \begin{equation} \small 
\label{eqt20} 
\alpha^{*}_{G(i),k} = \min \left\lbrace 1 - \sum_{j=1}^{i-1} \alpha_{j,k},  \max \{ \bar{m}_{i,k},0 \} \right\rbrace, \forall i\in\mathcal{M} 
\end{equation} 
\end{prop}

\noindent\emph{Sketch of the Proof}:
The objective function of $\mathcal{P}3$ can be decoupled into independent convex sub-problems that are concurrently solved by observing the virtual queues and channel states. The details are omitted due to the lack of space. \hfill $\blacksquare$

Following the solutions in Proposition~\ref{p1}, the matrix of the virtual queues $\mathcal{X}_{k}$ is updated at the end of each time $k$. The effectiveness of this method is validated in the next section.

\section{Simulation Results and Discussions}
\label{sec4}
In this section, the performance of the GPR based predictive control with the resource allocation method in Proposition 1 is investigated in an inverted-pendulum cart control task with $M=30$. Following~\cite{eisen2019control}, each cart system state is described by a four-dimensional vector, consisting of position and velocity of the cart along the  horizontal axis, and the angle and velocity of the pendulum. The control input is the horizontal force on the cart. By applying the zeroth-order hold on the continuous dynamics with a state sampling rate of $10$ ms, we obtain the following discrete-time linear dynamic matrices  of the inverted-pendulum on a cart system~\cite{eisen2019control}.  \begin{equation} \small 
\label{eqt21}
\mathbf{A}_{i} = \left[ \begin{array}{cccc}  
1 & 0       & 0& 0 \\
0 & 2.055     & -0.722 & 4.828 \\
0 & 0.023 & 0.91 & 0.037 \\
0 & 0.677 & -0.453 & 2.055 \\
\end{array} \right],  \mathbf{B}_{i} = \left[ \begin{array}{c}  
0.034  \\
0.168  \\
0.019  \\
0.105  \\
\end{array} \right]
\end{equation} 
To keep the system state close to zero, the feedback gain matrix $\mathbf{\Phi}_{i}$ is calculated according to the LQR controller in~\eqref{eqt3}. We perform a set of simulations scheduling the transmission  to control inverted pendulum system and perform scheduling using the proposed scheme which is compared to the round-robin scheduling scheme that is a type of time-triggered control system, i.e., the control system system transmit its state information in a predefined repeating order. In total with $100$ iterations, each simulation is run for $120$ seconds.
\begin{figure}
\label{fig2}
\centering
\includegraphics[width=0.9\columnwidth]{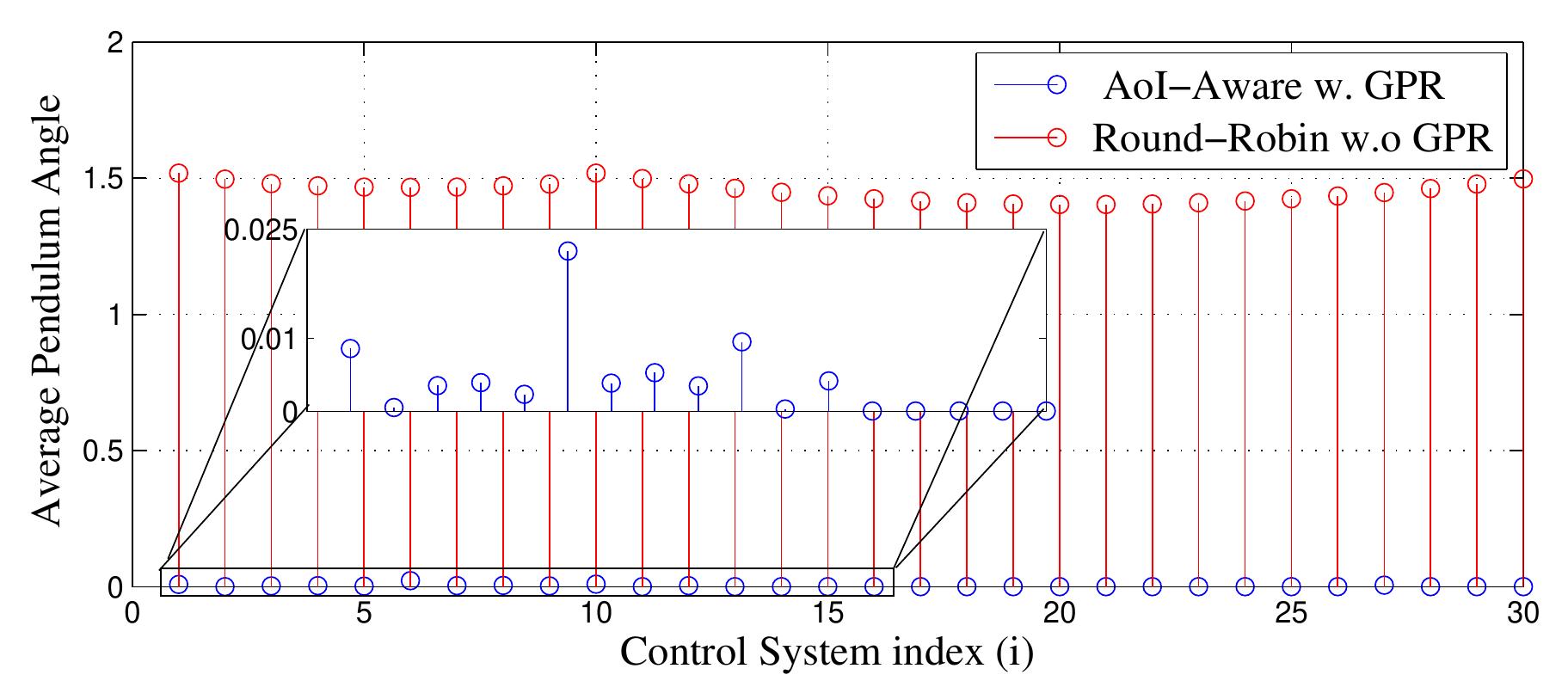}
\caption{Average pendulum angle to the vertical center, i.e., control error, with $M=30$ control systems. } \vspace{-5pt}
\end{figure}

Fig.~2 shows that the proposed method, i.e., \emph{AoI-aware scheduling with GPR}, achieves more than $75$x less average control errors for every control system compared to a \emph{round-robin baseline scheme without GPR}, in which the control error is measured by the average pendulum angle from the vertical center during $120$ seconds. 

The rationale behind this result is two-fold. As opposed to the baseline wherein only a single control system operates per each time due to the limited bandwidth, our proposed method allows all control systems to operate simultaneously even without receiving the current state, highlighting the effectiveness of GPR in enabling communication-efficient control. Furthermore, the proposed method applies AoI-aware scheduling that maintains the GPR prediction accuracy, thereby acheving control stability. To further validate this, Fig.~3 illustrates a single control system's AoI evolution over time. In sharp contrast to the baseline, after securing a certain amount of preceding states, GPR in the proposed method can locally predict the future states with high accuracy. Therefore, our method allows less frequent scheduling, observed by up to $67$\% higher peak AoI, advocating its communication efficiency.

A notable behaviour depicted in Fig.~3 is that the proposed method requires frequent scheduling at the early phase until $30$ seconds, within which the required amount of states is collected for accurate GPR prediction. During the same period, Fig.~4 exhibits the same tendency, in which each scheduling happens when the GPR prediction error $\text{Tr} \{ \mathbf{K}^{*}_{i,k} \}$ is higher than the MMSE estimation error $\text{Tr} \{ \mathbf{V}_{i,k} \}$, according to the discussion after \eqref{eqt9}. Based on this early-phase observation, sharing the GPR input DBs across different controllers (e.g., via broadcasting) could be an interesting topic for further improving communication efficiency. Another possible extension of this work is applying GPR not only to controllers but also to actuators.

\begin{figure}
\centering
\includegraphics[width=0.9\columnwidth]{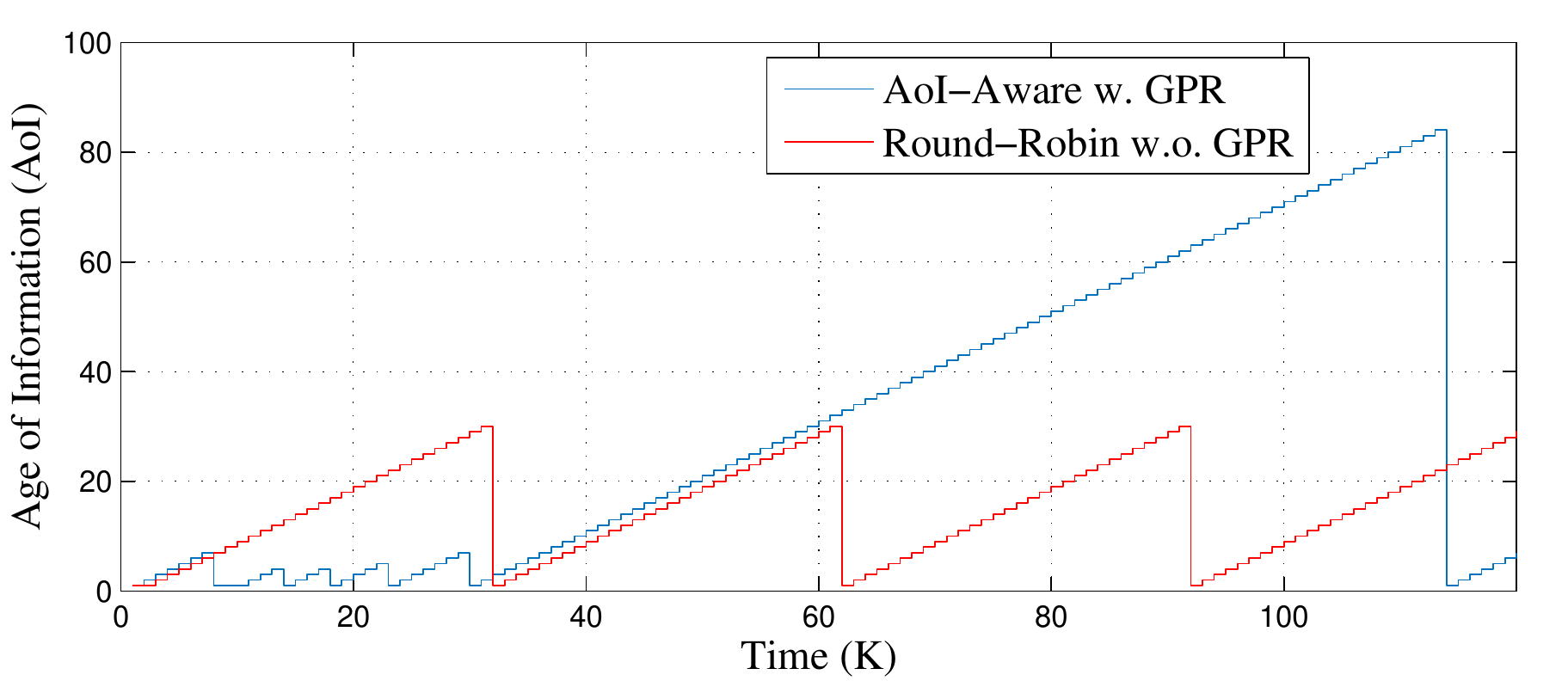} \label{fig3}
\vskip -15pt\caption{AoI of a randomly chosen control system with $M=30$. } \vspace{-5pt}
\end{figure}

\begin{figure}[t]
\label{fig4}
\centering
\includegraphics[width=0.9\columnwidth]{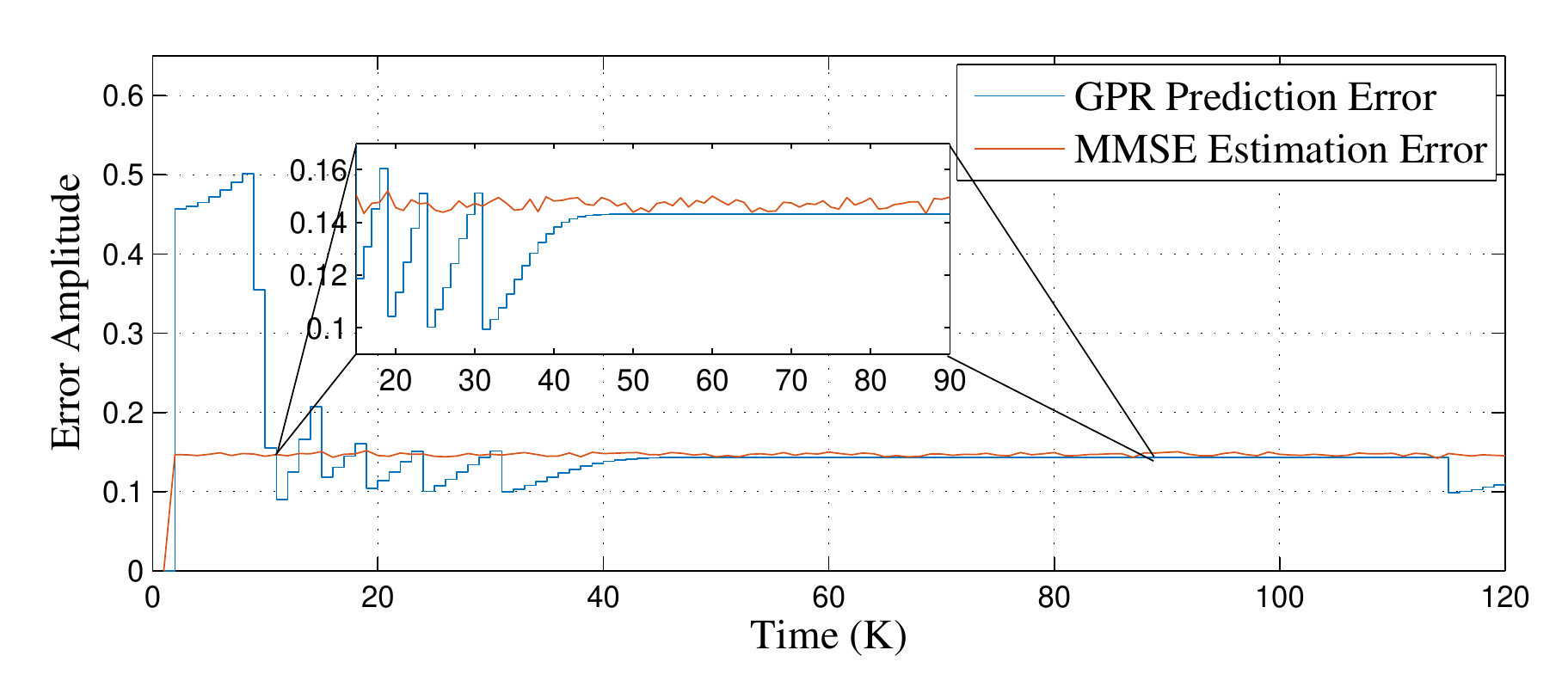}
\caption{GPR prediction and MMSE estimation errors of a randomly chosen control system with $M=30$.} \vspace{-5pt}
\end{figure}



 
\nocite{*}
\bibliographystyle{IEEEtran}
\bibliography{IEEEabrv,mybibfile}

\end{document}